\documentclass[%
manuscript,
amsmath,amssymb,amsthm,mathrsfs,
]{elsarticle}

\usepackage{graphicx}
\usepackage{dcolumn}
\usepackage{bm}
\usepackage[usenames]{color}
\usepackage[ 
margin=0.75in,
]{geometry}
\usepackage[normalem]{ulem}

\def\flinak{FLiNaK}
\def\degC{$^{\circ}$C}

\newcommand{\ra}[1]{\renewcommand{\arraystretch}{#1}}

\begin{document}
\begin{frontmatter}
	
\title{
	The Structure of Molten FLiNaK
}
	
\author[PhysAstro]{Benjamin A. Frandsen\corref{mycorrespondingauthor}}
\cortext[mycorrespondingauthor]{Corresponding author}
\ead{benfrandsen@byu.edu}

\author[ChemEng]{Stella D. Nickerson}

\author[ChemEng]{Austin D. Clark}
\author[ChemEng]{Andrew Solano}
\author[PhysAstro]{Raju Baral}
\author[ChemEng]{Jonathan Williams}
\author[ORNL]{J\"org Neuefeind}
\author[ChemEng]{Matthew Memmott}

\address[PhysAstro]{Department of Physics and Astronomy, Brigham Young University, Provo, Utah 84602, USA.}
\address[ChemEng]{Department of Chemical Engineering, Brigham Young University, Provo, Utah 84602, USA.}
\address[ORNL]{Neutron Scattering Division, Oak Ridge National Laboratory, Oak Ridge, Tennessee 37831, USA.}

\begin{abstract}
The structure of the molten salt (LiF)$_{0.465}$(NaF)$_{0.115}$(KF)$_{0.42}$ (FLiNaK), a potential coolant for molten salt nuclear reactors, has been studied by \textit{ab initio} molecular dynamics simulations and neutron total scattering experiments. We find that the salt retains well-defined short-range structural correlations out to approximately 9~\AA\ at typical reactor operating temperatures. The experimentally determined pair distribution function can be described with quantitative accuracy by the molecular dynamics simulations. These results indicate that the essential ionic interactions are properly captured by the simulations, providing a launching point for future studies of FLiNaK and other molten salts for nuclear reactor applications.\footnote{Notice: This manuscript has been coauthored by UT-Battelle, LLC under Contract No. DE-AC05-00OR22725 with the U.S. Department of Energy. The United States Government retains and the publisher, by accepting the article for publication, acknowledges that the United States Government retains a non-exclusive, paid-up, irrevocable, world-wide license to publish or reproduce the published form of this manuscript, or allow others to do so, for United States Government purposes. The Department of Energy will provide public access to these results of federally sponsored research in accordance with the DOE Public Access Plan (http://energy.gov/downloads/doe-public-access-plan).}
\end{abstract}


\end{frontmatter}

Molten salt reactors (MSRs) are a promising nuclear reactor concept in which fuel and/or fertile material are dissolved directly into a halide salt coolant. This has significant benefits over traditional light water reactors (LWRs) that are in operation today, including the capability of producing medical radioisotopes and electricity simultaneously in large amounts~\cite{epri;2015} and the possibility of reactor designs that prevent proliferation of weaponizable material, eliminate the risk of meltdown events, and avoid producing long-lived transuranic nuclear waste~\cite{ignat;ae12,mathi;nse09}.

The feasibility of molten salt thorium-fueled reactors and their potential for fuel breeding was validated through the Molten Salt Reactor Experiment (MSRE), which operated from 1965 to 1969 at Oak Ridge National Laboratory~\cite{aec;1972,rober;1971,betti;1972}. However, licensing commercial MSRs requires the validation of several additional fundamental chemistry concepts that were not accomplished in the MSRE, including chemical separation techniques, isotope extraction techniques, corrosion potential, ion-ion interaction potentials and subsequent impacts on thermophysical properties of the coolant, and the structure and speciation of the fission products in the salt. In addition, the MSRE project made no efforts to explore medical isotope production~\cite{brian;nse57}. Little work was done on MSRs for many years following the MSRE.

Growing interest in MSR technology in recent years has led to renewed research efforts on this topic~\cite{mathi;pne06,delpe;jfc09}. One of the most urgent issues is to establish the structure and speciation in the molten salt coolant with and without fission products~\cite{mscw;2017}, as this information plays a key role in developing transport and thermodynamic models. This is especially challenging because over 50 elements (fission products, fuel, and transuranic elements) are dissolved in the coolant at any given moment~\cite{souce;jnst05}. The first step toward this ambitious goal is to determine the structure of pure coolant salts without any fuel or fission products. Gaining insight into the speciation and complexation of the ions in the melt will provide a basis for understanding numerous basic properties such as viscosity, thermal conductivity, activity coefficients, and more, supporting the goal of developing reliable theoretical simulation models used for optimizing MSR systems.

In this work, we present pair distribution function (PDF) analysis of neutron total scattering data collected on the salt known as \flinak, a eutectic mixture of LiF, NaF, and KF that has been proposed as a promising salt coolant for use in MSR applications. PDF is an excellent tool for quantitatively probing the local structure of liquids, including molten salts~\cite{egami;b;utbp12,ender;advp80,rever;rpp86}. The PDF measurements reported here provide a rare opportunity to probe the structure of \flinak\ directly, marking an important development in the field. Further, \textit{ab initio} molecular dynamics (AIMD) simulations of \flinak\ were also performed, with the resulting structure predictions being compared directly to the PDF data.  We show that the AIMD structures reproduce the measured PDF data with quantitative accuracy. These results provide detailed information about the local ionic correlations in \flinak\ for the first time and demonstrate that the AIMD framework utilized here is reliable.

A pure, isolated sample of \flinak\ was produced for the neutron scattering experiment as follows. Individual components of LiF, NaF, and KF in a 46.5-11.5-42~mol~\% composition were dried in a vacuum oven for 6~hr at 300~\degC. These were then passed into an anaerobic, anhydrous argon-filled glovebox where they were combined and melted at 600~\degC\ in a nickel crucible. This melt was then subjected to argon sparging for 48~hr at a flow rate of 1~L/min to remove residual moisture and particulates. The molten salt mixture was then electrolyzed at a voltage of 1.5~V for 72~hr to remove soluble impurities. Lastly, the melt was filtered through an alumina filter with an average pore size of 2 microns to remove additional solid impurities and then allowed to cool and solidify.

Neutron total scattering experiments were performed on the NOMAD beamline~\cite{neuef;nimb12} at the Spallation Neutron Source using a vacuum furnace to access temperatures between 25~\degC\ and 1000~\degC. A 1.63~g sample of \flinak\ was crushed into a powder in an argon glove box near the beamline, loaded into a vanadium sample can, and placed in the beamline furnace chamber, which was subsequently evacuated. The closed sample can was in air for less than 30~s during this process. The total scattering structure function $S(Q)$ was obtained from the neutron time-of-flight data following standard data reduction protocols, including absolute normalization using a vanadium rod for reference. Scattering patterns obtained at temperatures below and above the melting point (454~\degC) were collected for a total integrated proton charge of 4~C and 12~C, respectively. The suppression of the scattered intensity at low momentum transfer $Q$ due to the large neutron absorption cross section of natural Li (arising from the $^6$Li isotope) was corrected empirically using a hydrogen-type Placzek correction fitted to the scattering pattern in the range of $1 < Q < 20$~\AA$^{-1}$.

The experimental pair distribution function (PDF) $g(r)$ was generated for each total scattering pattern according to the relationship~\cite{fisch;rpp06}
\begin{equation}\label{gr-sq}
	g(r)-1=\frac{1}{2\pi^2r\rho_0}\int_{0}^{Q_{\mathrm{max}}}Q\left[S(Q) - 1\right]\sin(Qr)\mathrm{d}Q,
\end{equation}
where $\rho_0$ is the number of atoms per unit volume and $Q_{\mathrm{max}} = 15$~\AA$^{-1}$, estimated to be the largest value of $Q$ for which meaningful scattering signal in the molten state remained larger than statistical noise in the data. For scattering patterns collected at temperatures below the melting point, the reduced PDF $G(r) = 4\pi r \rho_0 [g(r) - 1]$ was also generated with $Q_{\mathrm{max}}=20$~\AA$^{-1}$. Fits to $G(r)$ were performed using the program PDFgui~\cite{farro;jpcm07}.

AIMD simulations based on density functional theory (DFT) were performed using the CP2K software package~\cite{hutte;cms14}. Calculations utilized the Perdew-Burke-Ernzhof (PBE) exchange-correlation functional~\cite{perde;prl96}, a form of general gradient approximation (GGA). The Gaussian and plane waves method (GPW) was used, in which wave functions are represented by atom-centered Gaussian orbitals and the electron density is represented by plane waves~\cite{vande;cpc05}. Within this framework, all atoms were modeled using DZVP-MOLOPT-SR-GTH basis sets~\cite{goede;prb96} with core electrons treated with Goedecker-Teter-Hutter pseudopotentials~\cite{hutte;cms14,goede;prb96}. 

\begin{figure}
	\centering
	\includegraphics[width=80mm]{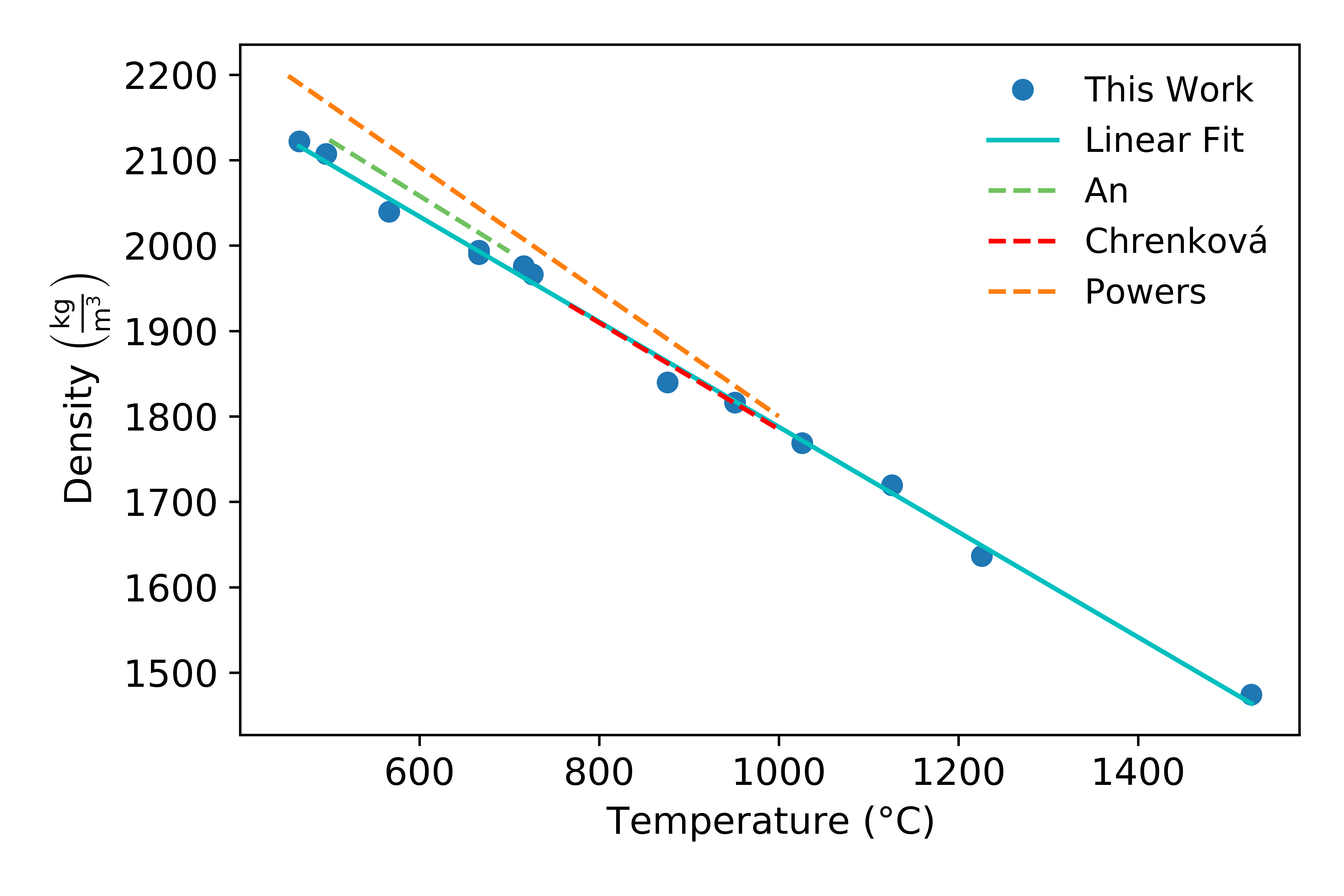}
	\caption{\label{fig:density} Density of FLiNaK determined from the AIMD simulations in the present work (blue circles, with the solid turquoise line showing a linear fit) compared to experimentally estimated values (dashed lines) from An \textit{et. al}~\cite{chren;jml03,power;nse63,an;acp17}, Chrenkov\'a \textit{et. al}~\cite{chren;jml03}, and Powers \textit{et. al}~\cite{power;nse63}. }
\end{figure}
Simulations were performed in the NPT ensemble, with pressure held at atmospheric pressure through the method of Ref.~\cite{marty;jcp94} and temperature held constant through the Nose-Hoover thermostat~\cite{hoove;pra85}. All simulations had a time step of 0.5~fs. The simulation cell was considered equilibrated if the average density over a period of 5~ps (10,000 steps) was within 2\% of the average density of the next 5~ps. The NPT ensemble is sensitive to relatively small errors in calculated energy, and so high energy cutoffs were required for predicted densities to converge: 2000~Ry for the plane-wave energy cutoff, and 120~Ry for the relative cutoff, which in the GPW method corresponds to the planewave cutoff of a reference grid on which Gaussians are mapped. The predicted densities were remarkably close to experiment. Fig.~\ref{fig:density} shows density versus temperature determined from MD simulations (each data point representing an equilibrated simulation cell containing 50 FLiNaK atoms) and compared to experimental results from the literature~\cite{chren;jml03,power;nse63,an;acp17}. An additional simulation using a cell with 100 atoms of FLiNaK at 667~$^{\circ}$C yielded an equilibrated density within 0.2\% of the expected value at that temperature from experimental correlation reported by Chrenkov\'a et al~\cite{chren;jml03} (well within the 2\% margin of error of that correlation). This simulation cell was used to calculate radial distribution functions for comparison with neutron scattering data. We note that including corrections for Van der Waals dispersion in the simulations resulted in a predicted density that differed from the experimental correlation by 4\%. It is known that the efficacy of Van der Waals dispersion corrections is dependent on the details of a given simulation, improving the accuracy in some cases but not in others (including the current study)~\cite{nam;jnm14}. Determining \textit{a priori} which combination of simulation methods is most legitimate is difficult, and comparison to experiment remains the best way of testing the validity of any particular set of methods. As such, dispersion corrections were omitted from all simulations used in this study.

\begin{figure}
	\centering
	\includegraphics[width=80mm]{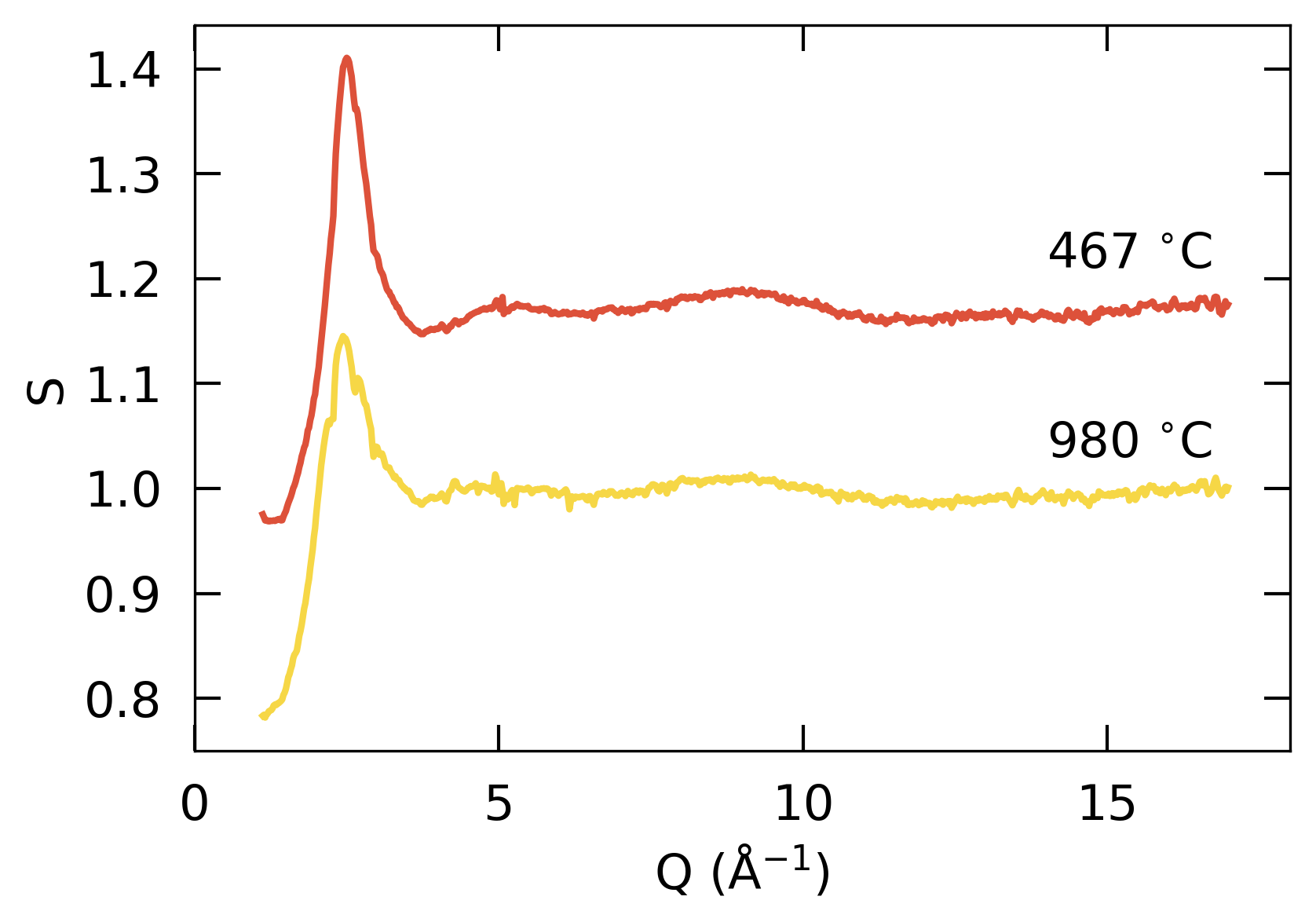}
	\caption{\label{fig:SQ} Total scattering structure function $S(Q)$ for molten \flinak\ at 467~\degC\ and 980~\degC, offset vertically for clarity.}
\end{figure}

We now present the results of our combined neutron total scattering and AIMD studies. In Fig.~\ref{fig:SQ}, we display the total scattering structure function $S(Q)$ for molten \flinak\ at 467~\degC\ and 980~\degC\ (the highest temperature attained during the experiment). The patterns are typical for liquids. The diffuse features in $S(Q)$ persisting out to approximately 12 - 15~\AA$^{-1}$ arise from well-defined short-range ionic correlations. The features in the scattering pattern become increasingly broad and weak as the temperature increases from 467~\degC\ to 980~\degC, indicating that the short-range correlations become less well-defined at higher temperatures. At lower temperature in the solid state (data not shown), sharp Bragg peaks are present, indicating crystalline order. Fits to the reduced PDF $G(r)$ for solid \flinak\ assuming a phase-separated mixture of the three binary salts LiF, NaF, and KF are significantly better than fits assuming a uniform distribution of the cations, indicating that solid \flinak\ has significant inhomogeneity.
	
A more intuitive view of the structure of \flinak\ can be gained by examining the real-space PDF patterns. Peaks in the PDF correspond to ion-ion pairs separated by the value of $r$ at which the peak is found, with the sign and size of the peak determined by the number of ion pairs separated by that distance and the product of the neutron scattering lengths of the corresponding ions.

\begin{figure}
	\centering
	\includegraphics[width=80mm]{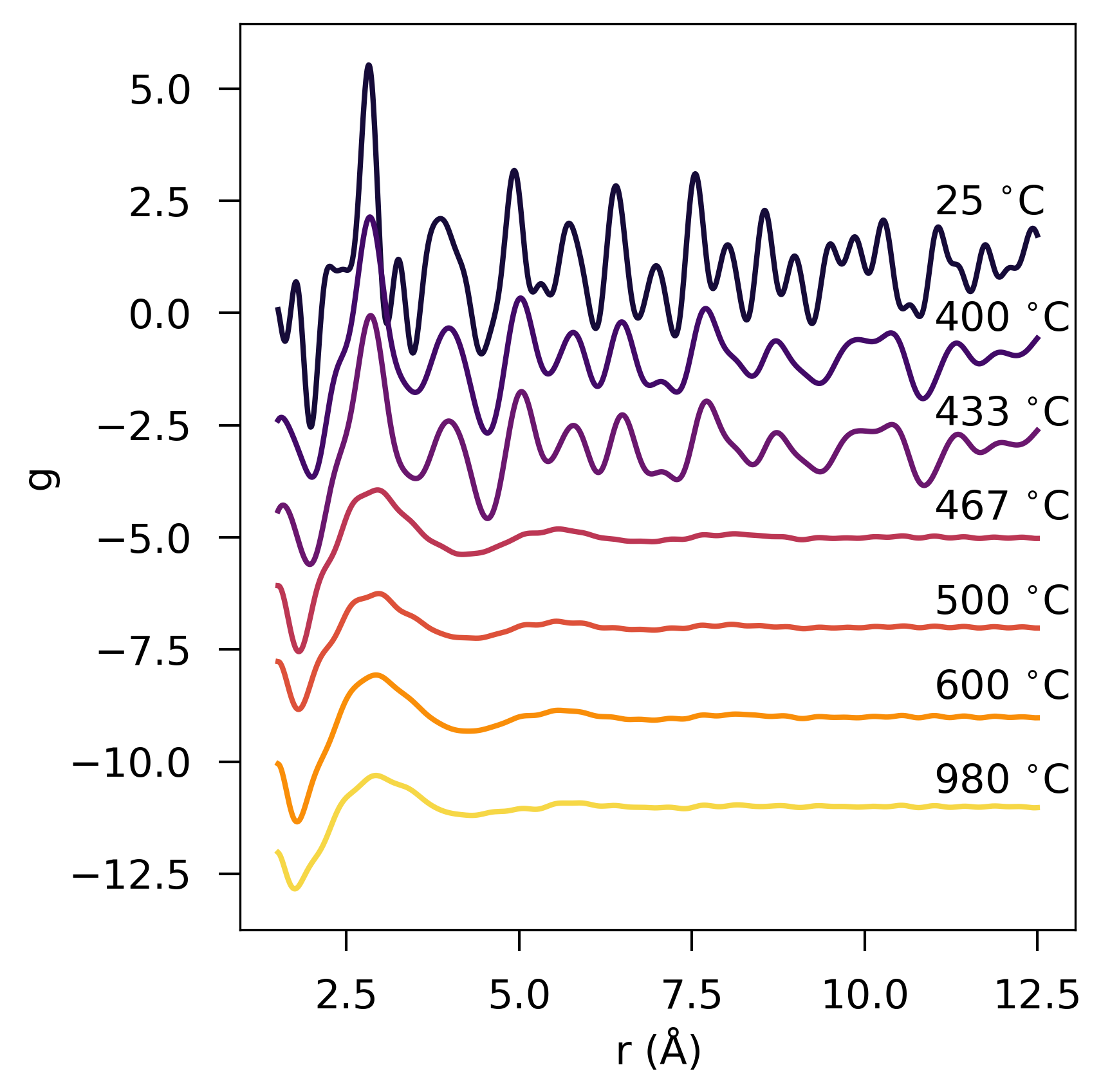}
	\caption{\label{fig:Grgr} The PDF $g(r)$ of \flinak\ at various temperatures below and above the melting point. The solid-liquid transition is clearly seen through the loss of well-defined features at high $r$ above the melting point. }
	
\end{figure}

To examine the evolution of the structure through the melting point, we plot the PDF $g(r)$ at several temperatures in Fig.~\ref{fig:Grgr}. At 25~\degC, well-defined features in the PDF persist over the entire displayed range, as expected for the long-range order in the solid state. The first peak in $g(r)$ at 25~\degC\ is negative and centered around $r = 2.0$~\AA, corresponding to the first nearest neighbor Li-F peak. Li has a negative scattering length and F has a positive scattering length, so all Li-F peaks are negative. The nearest-neighbor (NN) Na-F and K-F peaks are both positive due to the positive scattering lengths of Na and K, but they are less clearly visible in the pattern. From fits to the reduced PDF $G(r)$ at 25~\degC\ (not shown), the NN Na-F peak is expected to occur at 2.32~\AA. There is indeed a small feature in $g(r)$ at this position, but the low concentration of NaF greatly reduces its magnitude. The NN K-F peak is centered around 2.67~\AA, overlapping significantly with the first Li-Li and F-F peaks at 2.85~\AA\ from the LiF component. As the temperature is raised from 25~\degC\ to 433~\degC, the peaks in the PDF patterns become increasingly broad due to thermal vibration, but well-defined features persist over the full $r$ range.

The PDF patterns change dramatically in the molten state, as seen in Fig.~\ref{fig:Grgr} for the data collected at 467~\degC\ and above. A strong negative peak is still clearly visible at low $r$, indicating that the Li-F nearest neighbors are still well correlated. Interestingly, this first negative peak shifts from 2.0~\AA\ in the solid state to 1.8~\AA\ in the molten state, demonstrating that the strong attraction between Li$^+$ and F$^-$ results in a shorter average ionic bond in the molten state than in the solid state. A similar effect has been observed in other molten salts such as NaCl~\cite{biggi;jpcssp82}. On the basis of the AIMD simulations to be described later, we associate this shortening of the Li-F bond to a reduction of the coordination number from 6 in the solid state to 4 in the molten state.  The nearest-neighbor Na-F and K-F peaks are included in the broad positive peak observed between approximately 2~\AA\ and 3.5~\AA, along with contributions from various other ion pairs. Two additional weak and broad peaks centered around 5.5~\AA\ and 8~\AA\ can also be observed in Fig.~\ref{fig:Grgr} for the data collected at 467~\degC, 500~\degC, and 600~\degC, but beyond approximately 9~\AA, any remaining features in the PDF are comparable to the noise level. At 980~\degC, the broad peak centered on 8~\AA\ is not clearly resolvable, indicating that the non-random correlations on this length scale have been significantly weakened with increasing temperature.
	
We now turn to the AIMD simulations to extract more detailed information about the local structure and interactions in molten \flinak. After a given simulation converges, the ionic configurations can be averaged over a suitable period of time (40~ps in this case) and used to calculate the partial PDFs for each distinct type of ion-ion pair. These partial PDFs can then be summed together using the concentrations and scattering lengths as weights to generate the total PDF for comparison with the experimental PDF.

\begin{figure}
	\centering
	\includegraphics[width=80mm]{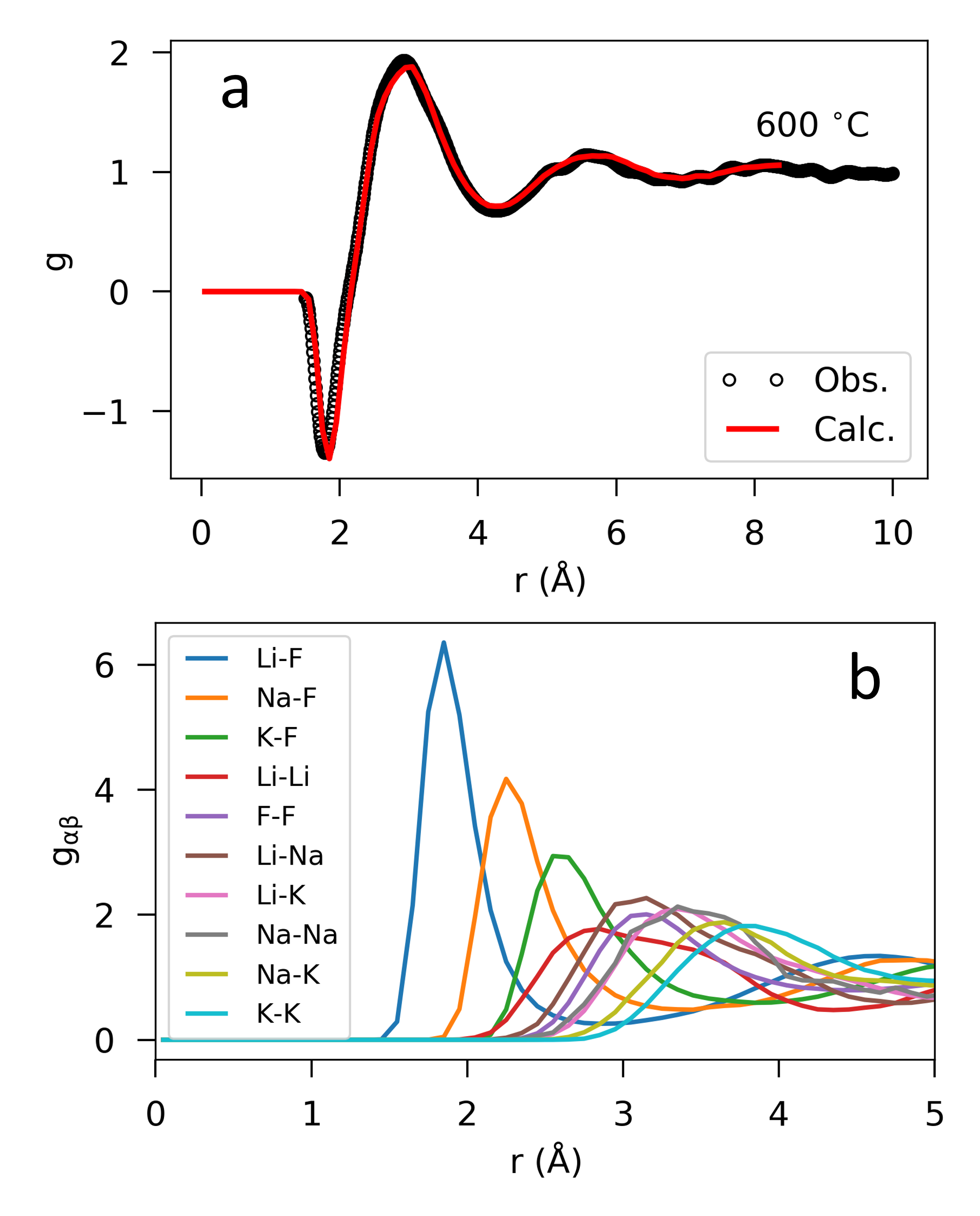}
	\caption{\label{fig:fit} (a) Observed PDF and calculated PDF determined from AIMD simulations of \flinak. The excellent agreement demonstrates that the simulations provide an accurate picture of the structure of moltent \flinak. (b) The partial PDFs $g_{\alpha \beta}$ extracted from the AIMD simulations, displayed over the first 5~\AA. Note that these partial PDFs have not been weighted by the concentrations or scattering lengths of the ions.}
	
\end{figure}

The calculated PDF determined from the AIMD simulation of \flinak\ at 667~\degC, a typical nuclear reactor operating temperature, is compared to the experimental PDF measured at 600~\degC\ in Fig.~\ref{fig:fit}(a). The experimental data and calculated PDF are shown as the black symbols and red curve, respectively. As seen in the figure, the calculated PDF provides an excellent match to the data over the first 8.35~\AA, confirming that the AIMD simulations yield a realistic model of the structure of molten \flinak. The simulation box size prevented us from calculating the PDF over larger distances, but this is not a significant limitation, since any observable features beyond $\sim$9~\AA\ are severely limited. In Fig.~\ref{fig:fit}(b), we show the partial PDFs $g_{\alpha \beta}$ determined from the simulations, corresponding to the correlations between each distinct type of ion-ion pair. Note that the partial PDFs have not been weighted by concentration or scattering length. The three most prominent peaks are Li-F, Na-F, and K-F, with maxima at approximately 1.84~\AA, 2.20~\AA, and 2.60~\AA, respectively. These peaks are all at slightly shorter distances than the nearest-neighbor cation-anion bond lengths in solid \flinak, with the Li-F peak being shifted to shorter $r$ most significantly. This is consistent with the observed shift in the negative Li-F peak across the melting point in Fig.~\ref{fig:Grgr}(b), demonstrating that the simulations accurately capture this effect. Subsequent partial PDFs are significantly broader, indicating that the corresponding correlations are more widely distributed in real space.

Nearest-neighbor (NN) peak distances and coordination numbers extracted from the AIMD simulations are shown for the most important ion pairs in Table~\ref{table:Coord}, along with the results from an earlier x-ray scattering experiment~\cite{igara;jcsft88}.
\begin{table}[ht]
	\ra{1.3}
	\caption{Nearest-neighbor (NN) peak distances and coordination numbers for various ion pairs determined from the AIMD simulations and an earlier x-ray scattering study~\cite{igara;jcsft88}.} 
	\centering 
	\begin{tabular}{l c c c c c} 
		\cline{2-6} 
		& \multicolumn{2}{c}{~~~NN peak (\AA)~~~} & \multicolumn{1}{c}{} & \multicolumn{2}{c}{Coordination Number} \\  
		\cline{2-6} 
		& AIMD & Exp.~\cite{igara;jcsft88} & ~~~~~ & AIMD & Exp.~\cite{igara;jcsft88} \\
		\hline
Li$^+$--F$^-$& 1.84 &1.83&  & 4.00&3.3 \\
Na$^+$--F$^-$& 2.20 &2.18&  & 5.42&3.8 \\
K$^+$--F$^-$& 2.60 &2.59&  & 7.12&4.0 \\
F$^-$--F$^-$& 3.10 &3.05&  & 11.2&8.9 \\
\hline

	\end{tabular}
	\label{table:Coord} 
\end{table}
The NN peak distances are in excellent quantitative agreement between the two studies. On the other hand, the coordination numbers show the same qualitative trend, but significant quantitative discrepancies exist. We note that the earlier x-ray study had much more limited sensitivity to the light elements in \flinak\ than the present neutron study, and that their method of determining the coordination numbers considered only the NN correlations of the three unlike-ion pairs and F$^-$--F$^-$ and K$^-$--K$^-$, ignoring all others. For these reasons, we consider the coordination numbers extracted from the present AIMD simulations to be more reliable.

The results presented here provide a crucial contact point between theory and experiment. The success of the AIMD simulations in describing the experimental PDF data demonstrates that we have an appropriate model for \flinak, which until now has been lacking a sound experimental basis. Detailed structural information from the simulations can therefore be extracted and used confidently for further theoretical and experimental work. Specifically, this work sets the stage for future studies that use the AIMD-predicted structure to explore solution chemistry and the solution complexes that form around fission products, providing direct insight into the behavior of the system under operating conditions in a molten salt reactor.

\textit{Acknowledgments} This research used resources at the Spallation Neutron Source, a DOE Office of Science User Facility operated by the Oak Ridge National Laboratory. BAF and RB gratefully acknowledge support from the College of Physical and Mathematical Sciences at Brigham Young University (BYU). SDN, ADC, and AS acknowledge support from the U.S. Department of Energy, Office of Nuclear Energy, Nuclear Energy University Program under Project 19-17413, CID: DE-NE0008870. 

\textit{Data Availability} Data will be made available upon request.
	

\end{document}